\documentclass[aps,pra,twocolumn,showpacs,superscriptaddress,floatfix]{revtex4-1}
\usepackage{graphicx,amsmath,amssymb}
\usepackage{bm}
\usepackage[usenames]{color}
\usepackage[dvipsnames]{xcolor}
\usepackage[colorlinks=true,linkcolor=blue,anchorcolor=blue,citecolor=blue,urlcolor=blue]{hyperref}
\begin{document}
	
	\title{A Kaleidoscope of Topological Structures in Dipolar Bose-Einstein Condensates with Weyl-Like Spin-Orbit Coupling in Anharmonic Trap
}
	
	\author{Yun Liu}
	\affiliation{School of Physical Science and Technology, Lanzhou University, Lanzhou 730000, China}
	\affiliation{Key Laboratory for Quantum Theory and Applications of MoE, Lanzhou Center for Theoretical Physics, Lanzhou University, Lanzhou 730000, China}
	
	\author{Zu-Jian Ying}
	\email{yingzj@lzu.edu.cn}
	\affiliation{School of Physical Science and Technology, Lanzhou University, Lanzhou 730000, China}
	\affiliation{Key Laboratory for Quantum Theory and Applications of MoE, Lanzhou Center for Theoretical Physics, Lanzhou University, Lanzhou 730000, China}

	\begin{abstract}
Dipole-dipole interaction (DDI) possesses characteristics different from the conventional isotropic s-wave interaction in Bose-Einstein condensates (BECs), the interplay of DDI with spin-orbit coupling (SOC) and rotation may induce novel quantum properties.
We systematically analyze the effects of the DDI, Weyl-like SOC, rotation and trap anharmonicity in the ground state of two-componen BECs. The interplay of these factors leads to a kaleidoscope of quantum states of quantum defects and quantum droplets in lattice, wheel and ring forms of distributions, with transitions of topology of density and a critical behavior in varying the parameters. We also show a bunch of exotic spin topological structures, including centric vortex surrounded by layers of spin flows, compound topological structure of edge defect, and various coexistence states of skyrmions with different topological charge. In particular, we find quarter skyrmions and other possible fractional skyrmions. Rashba-type SOC and Weyl-like SOC are compared as well. Our study implies that one can manipulate both the density topology and the spin topological structure via these tunable parameters in BECs. The abundant variations of the topological structures and particularly the revealed critical behavior may provide various quantum resources for potential applications in quantum metrology.
	\end{abstract}
	
	\maketitle
	
	\section{Introduction}

With the rapid development of studies on ultra-cold quantum gases, Bose-Einstein condensates (BECs)~\cite{cornell2002nobel} have become an ideal platform for exploring quantum phenomena of various interactions. Actullly BECs can accommodate not only isotropic short-range interactions~\cite{Krzysztof2000PRA} but also anisotropic long-range ones~\cite{Roccuzzo2019PRA,Su2024chaos,Li2022PRA,Halperin2023PRA,Zhang2024PRRes,Wang2024OE,Prasad2019PRA,Kazimierz2020PRL}. In particular, spinor BEC systems may also possess  different types of synthetic spin-orbit couplings (SOCs) which have a broad relevance to condensed matter~\cite{Krieger2024parallelSOC,JIANG2017PhysRepSkyrmions,ManchonNatMat-Rashba2015,Rashba1984,Dresselhaus1955},
nanosystems~\cite{Gentile2022NatElec,Ying2020PRR,Ying2017curvedSC,Ying2016Ellipse,Nagasawa2013Rings}, light-matter interactions~\cite{Ying-2021-AQT,Ying-gapped-top,Ying-Stark-top,Ying-Spin-Winding,Ying-JC-winding} and cold atoms~\cite{Liu2025Chaos,LiuYing02025exoticSOC2Ring,Liu2024oscillatory,Chen2024Chaos,GaoXianlong2023BECsolitons,GongMing2019BECSOC,LinRashbaBECExp2013Review,Li2012PRL,LinRashbaBECExp2011}.

Indeed, BECs can involve different interactions. On the one hand, the isotropic s-wave interaction between condensed atoms has been shown to explain most of the observed phenomena~\cite{Krzysztof2000PRA}. On the other hand, recent studies on atoms with large magnetic moments and polar molecules, such as Cr~\cite{Griesmaier2005PRL}, Er~\cite{Lu2011PRL} and Dy~\cite{Aikawa2012PRL}, have attracted significant interest in the physics of dipolar quantum gases, where dipole-dipole interactions (DDIs)~\cite{Roccuzzo2019PRA,Su2024chaos,Li2022PRA,Halperin2023PRA,Zhang2024PRRes,Wang2024OE,Prasad2019PRA,Kazimierz2020PRL} play a crucial role. Unlike the s-wave contact interaction, the DDI has long-range and anisotropic characteristics, which have a profound impact on the static properties, dynamic behavior and stability of the BEC~\cite{Su2024chaos,Li2022PRA,Adhikari2023PRA}. For instance, the DDI can give rise to novel quantum phases such as ferromagnetic superfluids~\cite{Lahaye2007Nature}, droplet crystals~\cite{Kadau2016Nature} and dipolar superfluids~\cite{Wenzel2018PRL} and unique phenomena like roton modes~\cite{Chomaz2018NaturePhy}. Furthermore, the DDI can be either attractive or repulsive and its tunability may lead to the emergence of new quantum phases.

The SOC is another area of broad interest~\cite{Krieger2024parallelSOC,JIANG2017PhysRepSkyrmions,ManchonNatMat-Rashba2015,Rashba1984,Dresselhaus1955,
Gentile2022NatElec,Ying2020PRR,Ying2017curvedSC,Ying2016Ellipse,Nagasawa2013Rings,
Ying-2021-AQT,Ying-gapped-top,Ying-Stark-top,Ying-Spin-Winding,Ying-JC-winding}, also receiving particular attention in recent investigations on quantum gas~\cite{Liu2025Chaos,
LiuYing02025exoticSOC2Ring,Liu2024oscillatory,Chen2024Chaos,GaoXianlong2023BECsolitons,GongMing2019BECSOC,LinRashbaBECExp2013Review,Li2012PRL,LinRashbaBECExp2011,
Saboo2024PRA,Yu2024NonlinearDynamics,Zhang2023Results,Huang2016NaturePhy,Kolkowitz2017Nature,Liu2024Results,Xu2022Chaos,Li2022Bessel}. The SOC describes the coupling between the spin and momentum of a particle and it is a key factor in determining the physical properties of BEC systems. The SOC can have different forms, including Dresselhaus-type SOC~\cite{Dresselhaus1955}, Rashba-type SOC~\cite{Rashba1984}, Weyl-type SOC~\cite{Anderson2012PRL,Krieger2024parallelSOC}, and their combinations\cite{Goldman2014Reports}. Under the influence of SOC, BECs exhibit a rich variety of topological structures, such as topological superfluid phase\cite{wu2016realizeSOC}, supersolid phase~\cite{Li2022PRA}, soliton excitations~\cite{Sakaguchi2017PRA,Gautam2018PRA,Wang2024NJP}, giant vortex~\cite{Wen2021Results,White2024PRA}, half-quantum vortex~\cite{Jung2023PRA,Ramachandhran2012PRA} and half-antiskyrmion fence~\cite{LiuYing02025exoticSOC2Ring}, further enhancing the potential of quantum gas studies. Particularly, the combination of the DDI and SOC is expected to induce more complex quantum states and novel phase transitions, laying a crucial theoretical and experimental foundation for exploring intriguing phenomena in dipolar BECs.

It should be reminded that, in real physical systems, ultra-cold dipolar BECs are confined in external potential traps, so their properties are closely related to the profile and geometry of the potential~\cite{Jiao2022Frontiers,Mujal2020PRA,Liu2025Chaos,Shirley2014PRL,Li2024Merging,LiuYing02025exoticSOC2Ring,Wang2017toroidal,Thomas2017PRA,Kunimi2019PRA,Zhang2022toroidal,Liu2022toroidal,Roussou2018NJP}. Although there are studies on special potential such as optical lattices~\cite{Chen2024Chaos,Sun2016Scientific,Salerno2016PRA}, double-well potentials~\cite{Li2024Merging} and ring-shaped traps~\cite{Wang2017toroidal,Thomas2017PRA,Kunimi2019PRA,Zhang2022toroidal,Liu2022toroidal,Roussou2018NJP}, a most primarily used trap is the harmonic potential~\cite{Jiao2022Frontiers,Mujal2020PRA,Liu2025Chaos,Shirley2014PRL} while the effect of anharmonicity in real systems needs to be examined. Furthermore, rotation provides a versatile platform for engineering exotic quantum phases in ultracold gases. While in conventional BECs it gives rise to quantized vortices~\cite{abo2001observation} and vortex lattices~\cite{Fetter2009RevModPhys}, its role in dipolar BECs with SOC particularly within anharmonic traps remains unexplored.

Taking into account of all these key factors, i.e. the DDI, SOC, potential and rotation, a thorough investigation on dipolar two-component BECs with both Weyl-like SOC and rotation in anharmonic traps is lacking. One may wonder what novel quantum phases and intriguing physical properties might emerge in such systems.

In this work, we investigate the topological structures in the density distributions and spin textures of rotating two-dimensional (2D) spin-1/2 dipolar BECs with Weyl-like SOC in an anharmonic trap. Using the imaginary time evolution method, we obtain the system ground states and systematically analyze the effects of the DDI, Weyl-like SOC, rotation and trap anharmonicity on the density distribution and sin texture. We reveal a kaleidoscope of exotic quantum states, with transitions of density topology,  critical behavior around a revealed collapse point and a bunch of coexisting states of spin structures with integer and fractional topological charges. Our study indicates the possibility for manipulation of both the density topology and the spin topological structure in BECs and potential quantum resources for applications in quantum metrology.

This paper is organized as follows.
Sec.~\ref{Model} introduces the model and the numerical method.
Sec.~\ref{Sect-Effects} addresses the effects of the DDI, Weyl-like SOC, rotation and trap anharmonicity respectively on the density distributions. Transitions of density topology and critical behavior are noticed.
Sec.~\ref{Sect-SpinTexture} shows the spin textures of different states, revealing various exotic topological spin structures and coexistence states of different topological charges. Rashba-type SOC and Weyl-like SOC are compared.
Sec.~\ref{Conclusion} summarizes the main results.

	\section{Model and method}
	\label{Model}
	We consider a quasi-2D system of two-component rotating BECs with DDI and Weyl-like SOC in an anharmonic trap. The system is composed of atoms possessing magnetic dipole moments, referred to as component $1$, along with nonmagnetic atoms, designated as component $2$. In the framework of the mean-field theory, the ground state and the dynamics of the system can be described by the following nonlinear coupled Gross-Pitaevskii (GP) equations:
	\begin{eqnarray}
&& i\hbar \frac{\partial \psi_1}{\partial t} = \left[ -\frac{\hbar^2 \nabla^2}{2m} + V(\mathbf{r}) + g_{11} |\psi_1|^2 + g_{12} |\psi_2|^2 - \Omega L_z \right. \notag \\
&& \left. + \int U_{dd}({\bf r}-{\bf r}^{\prime}) |\psi_1(\mathbf{r}^\prime, t)|^2  d\mathbf{r}^\prime \right] \psi_1	- k (\partial_y + i \partial_x) \psi_2 \notag \\
&& i\hbar \frac{\partial \psi_2}{\partial t} = \left[ -\frac{\hbar^2 \nabla^2}{2m} + V(\mathbf{r}) + g_{21} |\psi_1|^2 + g_{22} |\psi_2|^2 - \Omega L_z \right] \psi_2 \notag \\
&& + k ( \partial_y - i \partial_x) \psi_1  \label{Eq-GP}
	\end{eqnarray}
	where $\psi_{j}(j=1,2)$  is two-component wave function and satisfies the normalization condition $\int dxdy(|\psi_1|^2+|\psi_2|^2)=N$, $N$ represents the total number of particles, m is the atomic mass, $\Omega$ is the effective rotation frequency and $L_z=-i \hbar(x\partial_y-y\partial_x)$ is the $z$-component of the orbital angular momentum operator. The coefficients $g_{j j}=2\sqrt{2\pi}a_{j}\hbar^{2}/m a_{z}(j=1,2)$ and $g_{12 }=g_{21}=2\sqrt{2\pi}a_{12}\hbar^{2}/m a_{z}$ denote the intra- and inter-species contact interactions, $a_j(j=1,2)$ and $a_{12}$ are related to the s-wave scattering lengths between the intra- and inter-component atoms, respectively. $a_{z}$
is the oscillation length in the z direction. $k$ characterizes the strength of Weyl-like SOC. The external trap $V$ is the combined potential of a 2D harmonic and a 2D Gaussian trap, which can be expressed as:
	\begin{eqnarray}
		V=\frac{1}{2}m\omega_{\bot}^{2}r^{2}+V_{0}e^{\frac{-2r^{2}}{\omega_{0}^{2}}}
	\end{eqnarray}
	here, $\omega_{\perp}$ is the radial trapping frequency, $r=(x,y)$ is the two-dimensional coordinate with $r^2=x^2+y^2$. $V_0$ and $\omega_0$ denote the intensity and waist of the Gaussian beam, respectively.
	
	The long-range nonlocal DDI potential can be described by
	\begin{eqnarray}
		U_{d d}({\bf r}-{\bf r}^{\prime})=C_{dd} \frac{1-3\cos^{2}\theta}{\left|{\bf r}-{\bf r}^{\prime}\right|^{3}},
	\end{eqnarray}
	where $C_{dd}=\mu_0\mu^2/4 \pi$ is the strength of the DDI for magnetic dipoles, $\mu_0$ and $\mu$ are the magnetic permeability of vacuum and the magnetic dipole moment of the atoms, respectively. $\cos\theta=\mathbf{n}\cdot\mathbf{r}/\vert\mathbf{r}\vert$ with $\theta$ being the angle between the polarization direction $\mathbf{n}$ and the relative position $\mathbf{r}$ of the atoms.

To facilitate numerical computations and simulations, we can reformulate the GP equation in a dimensionless representation with the renormalized contact interactions
\begin{eqnarray}
    \beta_{j j}=g_{j}N m/\hbar^{2}(j=1,2),
	\quad \beta_{12}=g_{12}N m/\hbar^{2},
\end{eqnarray}
and the DDI tuned by
\begin{equation}
\varepsilon_{dd}=a_{dd}/a_{s}=\mu_0\mu^2m/3\sqrt{2\pi}\hbar^2a_{z}/a_{s},
\end{equation}
via the relation $a_{dd}=4\pi C_{dd}/3\sqrt{2\pi}a_{z}$. We leave the detail of the equation reformulation and numerical treatments in the Appendix \ref{Apendix-method}.
	
	\begin{figure*}[t]
		\includegraphics[width=2\columnwidth]{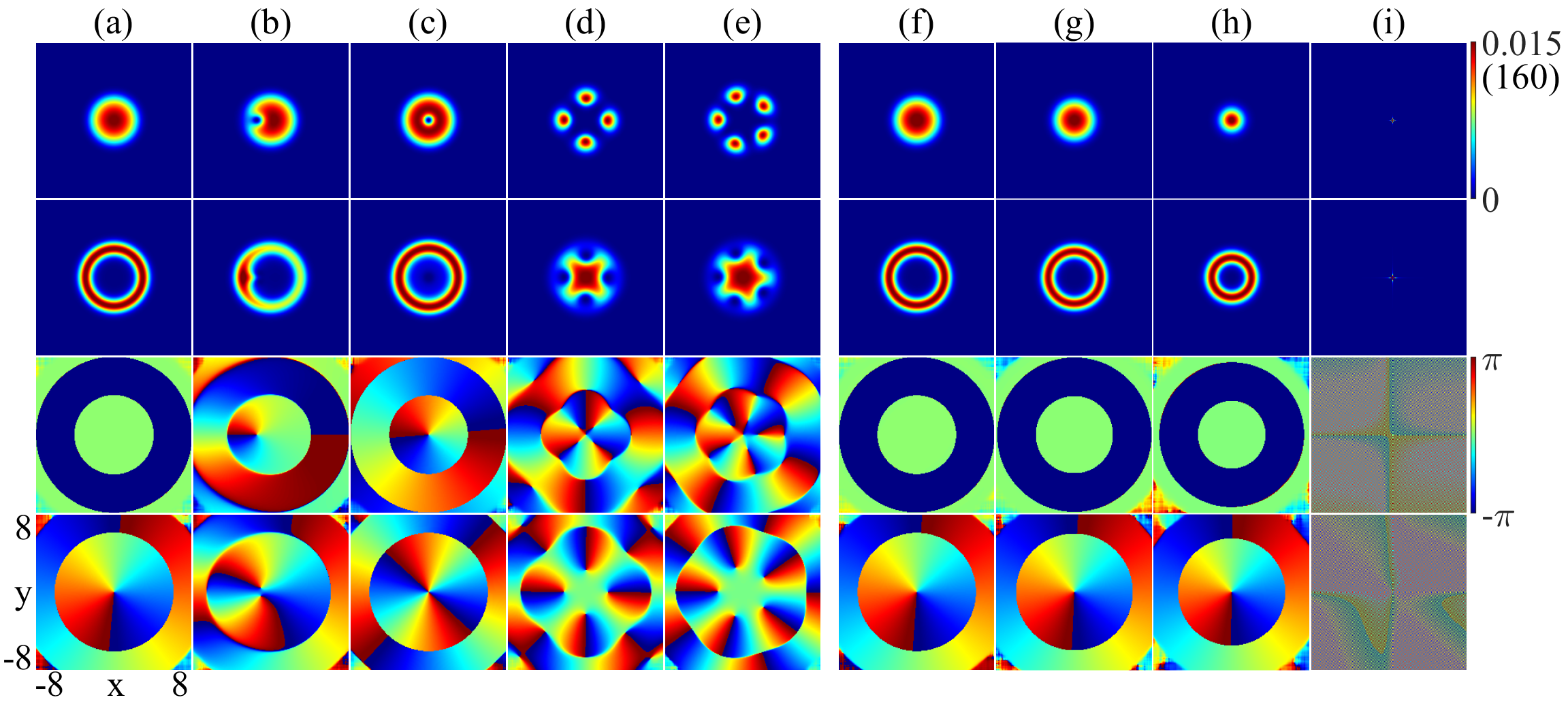}
		\caption{Effect of varying the DDI strength $\varepsilon_{dd}$ on the ground-state density distribution with fixed SOC strength $k=0.8$ and rotation frequency $\Omega=0.3$.
(a) $\varepsilon_{dd}=0.0$,
(b) $\varepsilon_{dd}=0.2$,
(c) $\varepsilon_{dd}=0.5$,
(d) $\varepsilon_{dd}=1.0$.
(e) $\varepsilon_{dd}=2.5$.
(f) $\varepsilon_{dd}=-0.2$.
(g) $\varepsilon_{dd}=-0.5$.
(h) $\varepsilon_{dd}=-1.0$.
(i) $\varepsilon_{dd}=-1.5$.
Other parameters are $\beta_{11}=50,\beta_{22}=80,\beta_{12}=200$. The upper two rows represent the density distribution $|\psi_1|^2,|\psi_2|^2$, and the lower two rows denote the phase distribution arg$(\psi_1)$,arg$(\psi_2)$ respectively. The color bar in the first row denotes the density range with the maximum value $0.015$ for (a)-(h) and $160$ for (i).}
		\label{Fig-DDI}
	\end{figure*}

\section{The effects of DDI, SOC, rotation and potential anharmonicity in their interplay}\label{Sect-Effects}

	\subsection{The effect of DDI}

	\subsubsection{Repulsive DDI: quantum defects and droplets in phase separation and transitions of density topology}
	
	We first investigate the effect of varying the DDI strength $\varepsilon_{dd}$ with fixed SOC strength $k$ and rotation frequency $\Omega$. Fig.~\ref{Fig-DDI} shows the density and phase distribution, with the rows from top to bottom representing $|\psi_1|^2$, $|\psi_2|^2$, $\arg(\psi_1)$ and $\arg(\psi_2)$, respectively.

In the absence of DDI as in Fig.~\ref{Fig-DDI}(a), the density of the first component ($\psi_1$) is distributed like a disk, while the second component ($\psi_2$) forms a density ring surrounding the first component. Such a density distribution results from energy reduction of the contact interaction in rotation. In this case the phase of the fist component is a homogenous thick ring while the second component accumulates $2\pi$ of phase in going around the origin. In this situation there is a vortex at the center (the origin). We leave the discussion of the spin texture in Section \ref{Sect-SpinTexture}.

At a dipolar interaction strength of $\varepsilon_{dd}=0.2$ as in Fig.~\ref{Fig-DDI}(b), a dent defect appears at the edge of the density disk of the first component, while the density ring of the second component gets a bulge at the defect. This defect is accompanied with a compound skyrmion.

At $\varepsilon_{dd}=0.5$ in Fig.~\ref{Fig-DDI}(c), the previous edge defect of the first component moves to the center of the density disk to form a thick density ring, while the second component recovers the regular shape of density ring, additionally with a low-density disk inside.  In such a situation the center defect also hosts a vortex.

As the DDI strength increases more to be $\varepsilon_{dd}=1.5$ in Fig.~\ref{Fig-DDI}(d) and $\varepsilon_{dd}=2.5$ in Fig.~\ref{Fig-DDI}(e), the first component forms four quantum droplets in an arrangement of square-corner orientations, while the second component shows a cross profile in the main density shape.

When the DDI strength reaches $\varepsilon_{dd}=2.5$ in Fig.~\ref{Fig-DDI}(e), the quantum droplets of the first component increases in number and are distributed in pentagonal orientations, while the second component forms a pentagram shape in the main part of density.

The spin distributions in the cases of Fig.~\ref{Fig-DDI}(d) and \ref{Fig-DDI}(e) generate square-corner and pentagonal vortex lattices, respectively. We will illustrate the figures of vortex later on in Section \ref{Sect-SpinTexture}.

Besides the variations in the number and the positions of these topological defects, it should be noted that actually there are also transitions of topology in the density distribution. Indeed, in the above increase of the DDI strength, the system remains in a phase-separated state in the sense that the two components tend to stay apart. The first component turns out to be embedded in density holes of the second component. Note that there is only one big density hole in the cases of Figs.~\ref{Fig-DDI}(a)-\ref{Fig-DDI}(c), while more small density holes emerge in Figs.~\ref{Fig-DDI}(d) and \ref{Fig-DDI}(e). In this sense the density topology is changed as the hole number varies.

We also notice that in the phase distributions [$\arg(\psi_1)$ and $\arg(\psi_2)$] the phase boundaries of the two components have different radii, which actually separate spin flows in opposite directions, as we shall present in Section \ref{Sect-SpinTexture}.

\begin{figure}[t]
\centering
\includegraphics[width=0.9\columnwidth]{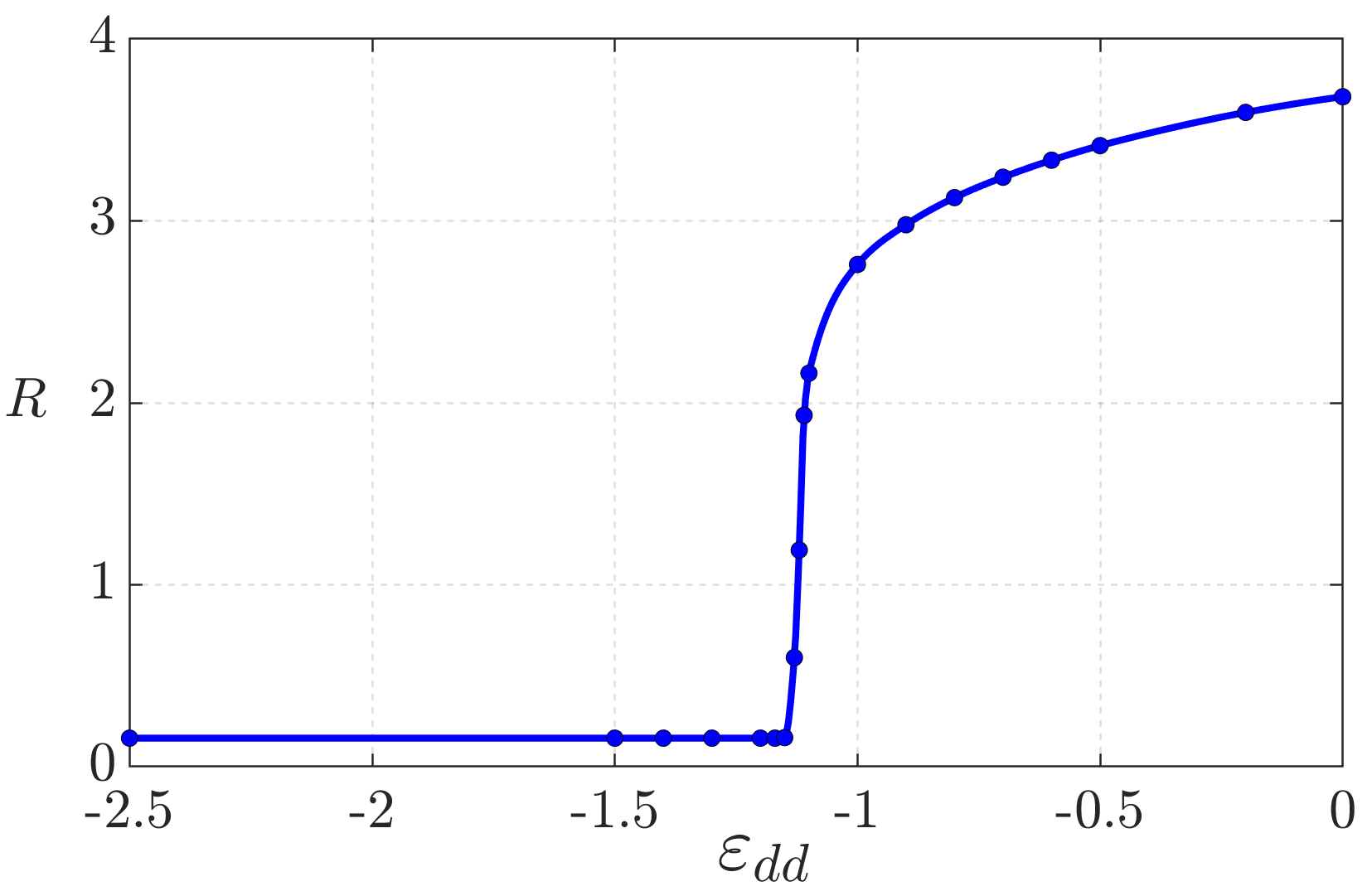}
\caption{Collapse transition and critical behavior in attractive DDI: Density radius of the second component $R$ versus the DDI strength $\varepsilon_{dd}$. Here $k=0.8$, $\Omega=0.3$, $\beta_{11}=50,\beta_{22}=80,\beta_{12}=200$ as in Fig.~\ref{Fig-DDI}}
\label{Fig-Critical-r}
\end{figure}

\subsubsection{Attractive DDI: critical transition of condensate radius}

We find a critical point in attractive DDI. Figs.~\ref{Fig-DDI}(f)-\ref{Fig-DDI}(i) show some examples of density and phase distributions in attractive DDI. From Figs.~\ref{Fig-DDI}(f)-\ref{Fig-DDI}(h) we see that the two components remain in a phase separated state with a disk-like density distribution in the first component and a ring shape in the second component, while the radii is deceasing when the DDI strength is increased. When the attractive DDI exceeds some critical point, both the two components collapse into spike-like distributions, as in \ref{Fig-DDI}(i) where the densities appear like sharp points around the origin.

The variation before the critical point is more continually shown in Fig.~\ref{Fig-Critical-r} by the evolution of the ring radius of the second component. The radius evolution turns out to be critical until the collapse point. Such a critical behavior manifests a fast variation in the ground state, which can provide sensitivity resource for critical quantum metrology~\cite{RamsPRX2018,Garbe2020,Montenegro2021-Metrology,Ilias2022-Metrology,Ying2022-Metrology,Hotter2024-Metrology,Ying-Topo-JC-nonHermitian-Fisher,
Ying-g2hz-QFI-2024,Ying-g1g2hz-QFI-2025,Ying2025g2A4,Ying2025TwoPhotonStark}.
	
	\begin{figure*}
		\includegraphics[width=0.8\textwidth]{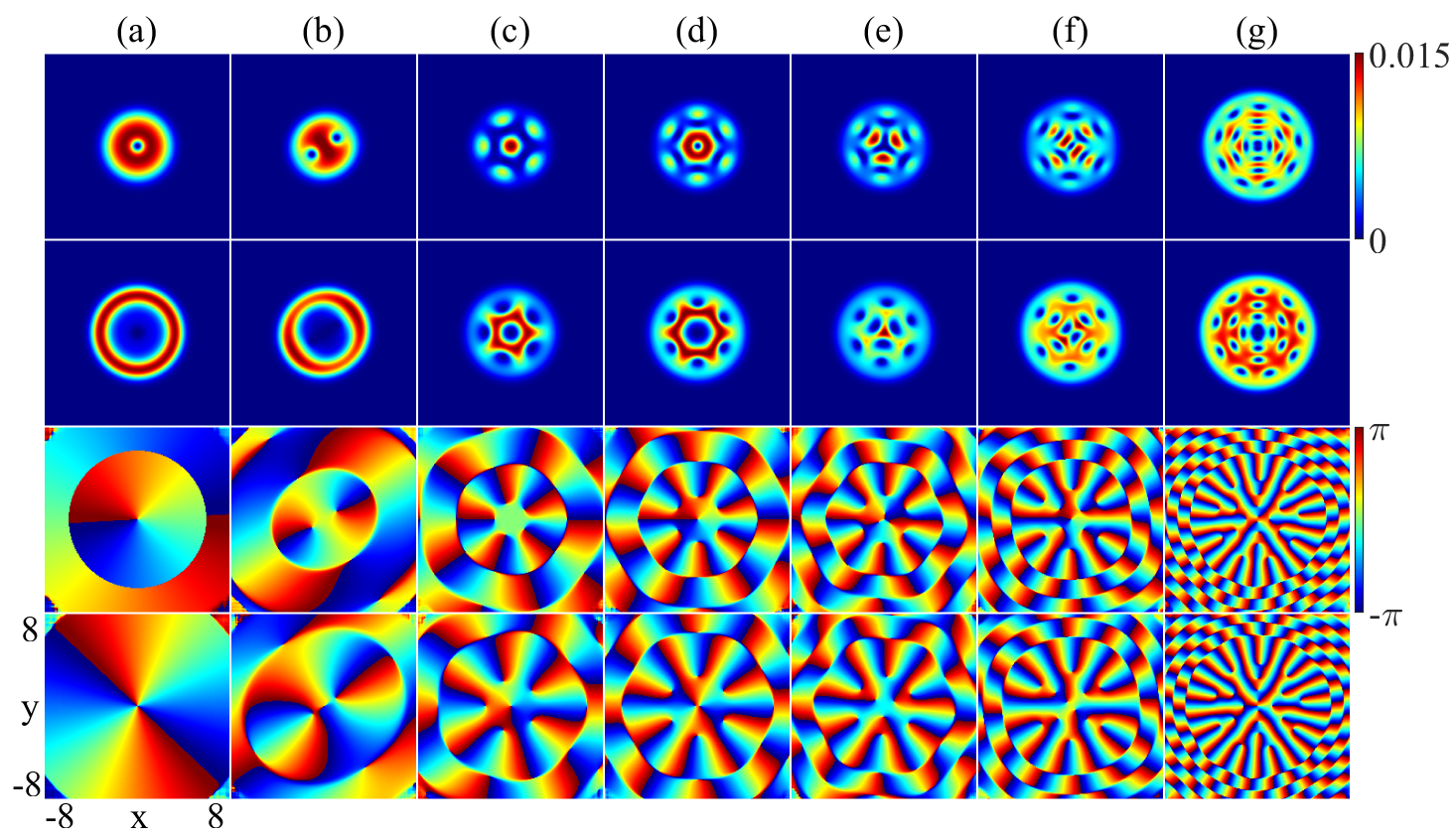}
		\caption{Effect of varying the SOC strength $k$ on the ground-state density distribution in fixed DDI $\varepsilon_{dd}=0.5$ and rotation frequency $\Omega=0.4$. (a) $k=0.4$, (b) $k=0.7$, (c) $k=1$, (d) $k=1.2$, (e) $k=1.6$, (f) $k=2$, (g) $k=3.5$. Other parameters are $\beta_{11}=50,\beta_{22}=80,\beta_{12}=200$. The first two rows represent the density distribution, and the last two rows represent the phase distribution. respectively.}
		\label{Fig-SOC}
	\end{figure*}	
		\begin{figure*}
			\includegraphics[width=0.8\textwidth]{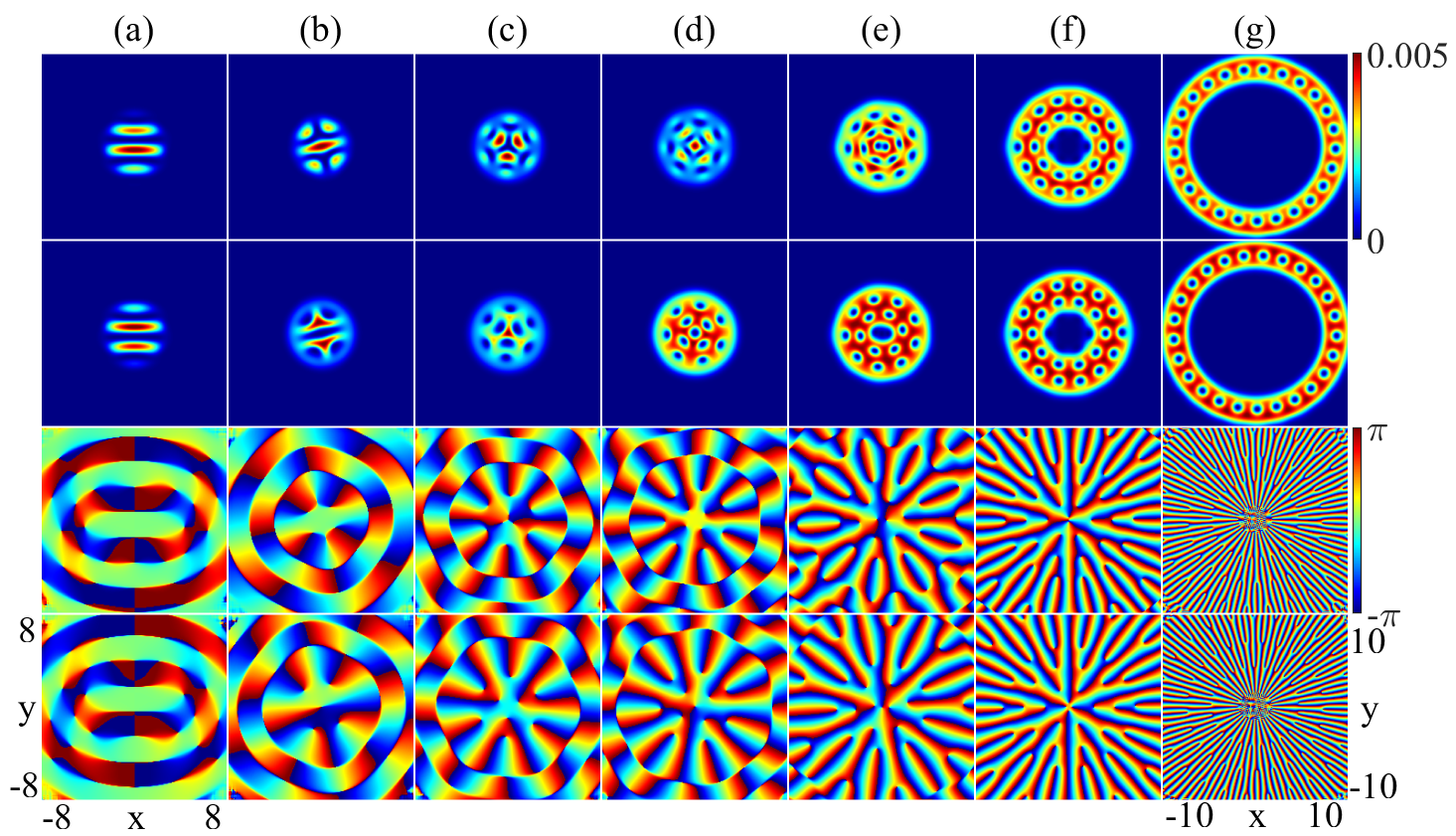}
			\caption{Effect of rotation on the ground-state density distribution in fixed DDI $\varepsilon_{dd}=0.5$ and SOC $k=1.5$. (a) $\Omega=0$, (b) $\Omega=0.2$, (c) $\Omega=0.4$, (d) $\Omega=0.5$, (e) $\Omega=0.6$, (f) $\Omega=0.7$, (g) $\Omega=0.83$. Other parameters are $\beta_{11}=50,\beta_{22}=80,\beta_{12}=200$. The first two rows represent the density distribution, and the last two rows represent the phase distribution. respectively.}
			\label{Fig-Rotation}
		\end{figure*}

\subsection{The effect of SOC}

We then check the effect of the SOC with given DDI and rotation frequency. Fig.~\ref{Fig-SOC} shows the density and phase distributions in varying the SOC strength $k$, with the rows from top to bottom also representing $|\psi_1|^2$, $|\psi_2|^2$, $\arg(\psi_1)$ and $\arg(\psi_2)$.

With a SOC strength $k=0.4$ in Fig.~\ref{Fig-SOC}(a) the density distribution is similar to the case in Fig.~\ref{Fig-DDI}(c). As the SOC strength increases to $k=0.7$, the center defect of the first component moves the edge of density disk, which is an opposite variation with respect to that of the DDI from Fig.~\ref{Fig-DDI}(b) to \ref{Fig-DDI}(c), except that the defect number is two here.

At $k=1.0$ in Fig.~\ref{Fig-SOC}(c), the quantum droplets in pentagonal orientations emerge in the first component, while the outer ring structure of the second component is converted into a pentagram star. This density distribution is similar to that in Fig.~\ref{Fig-DDI}(e) but with an additional quantum droplet at the center and an extra hole in the pentagram density, respectively in the two components.

At $k=1.2$ in Fig.~\ref{Fig-SOC}(d), the first component has six outer quantum droplets and an inner hexagon, while the 2nd component has six outer holes and an inner hexagram.

At $k=1.6$ in Fig.~\ref{Fig-SOC}(e), the inner hexagon of the first component is broken into three pieces and an inner Hexagon, while the inner hexagram turns to three holes surrounding a triangle density at the origin, looking like a ghost face.

Enhancing the SOC more leads to larger sizes of the density wheels with more layers of defects, as in Figs.\ref{Fig-SOC}(f) and \ref{Fig-SOC}(g). As will be addressed in Section \ref{Sect-SpinTexture}, these defects accommodate different skyrmions. We also see that the miscibility of the two components increases, in comparison with the phase-separated distribution in weaker SOC. Correspondingly in the phase distribution the layer number increases in the radial direction and more periods of $2\pi$ phase accumulations are acquired in the circular direction surrounding the origin, in the enhancements of the SOC.

\subsection{The effect of rotation}\label{Sect-Rotation}

We now analyze the effect of rotation by varying the rotation frequency $\Omega$ with fixed DDI and SOC strengths.

In the absence of rotation, as shown in Fig.~\ref{Fig-Rotation}(a), the density distributions of the two components exhibit a horizontal stripe pattern, the stripes of the two components are displaced in a phase-separated state. In tis case there is no vortex in the main density region but one can find the formation of ghost vortices at the periphery of the two components, which carry neither energy nor angular momentum~\cite{Wen2010double-well}.

At $\Omega=0.2$, as illustrated in Fig.~\ref{Fig-Rotation}(b), the first component displays a mixed state of quantum droplets and stripe state, while the second component exhibits a mixed state of density holes and stripes with protrusion. At $\Omega=0.4$ in Fig.~\ref{Fig-Rotation}(c), the two components form two layers of quantum droplets and holes, with a ghost face shape similar to Fig.~\ref{Fig-SOC}(e).

With the increase of the rotation frequency, the miscibility of the
two components increases and the quantum droplets disappear, the density wheel expands in radius size, and both the number and layers of vortices gradually grow. Such an evolution tendency results in a multi-layered structure of defects, as depicted in Figs.~\ref{Fig-Rotation}(d)-\ref{Fig-Rotation}(e) with $\Omega=0.5,0.6,0.7,0.83$. When the rotation frequency exceeds a critical value around $\Omega=0.6$, the density distribution transits from a defect disk to a defect ring, as shown in Figs.~\ref{Fig-Rotation}(e)-\ref{Fig-Rotation}(g).

	\begin{figure}
		\includegraphics[width=0.5\textwidth]{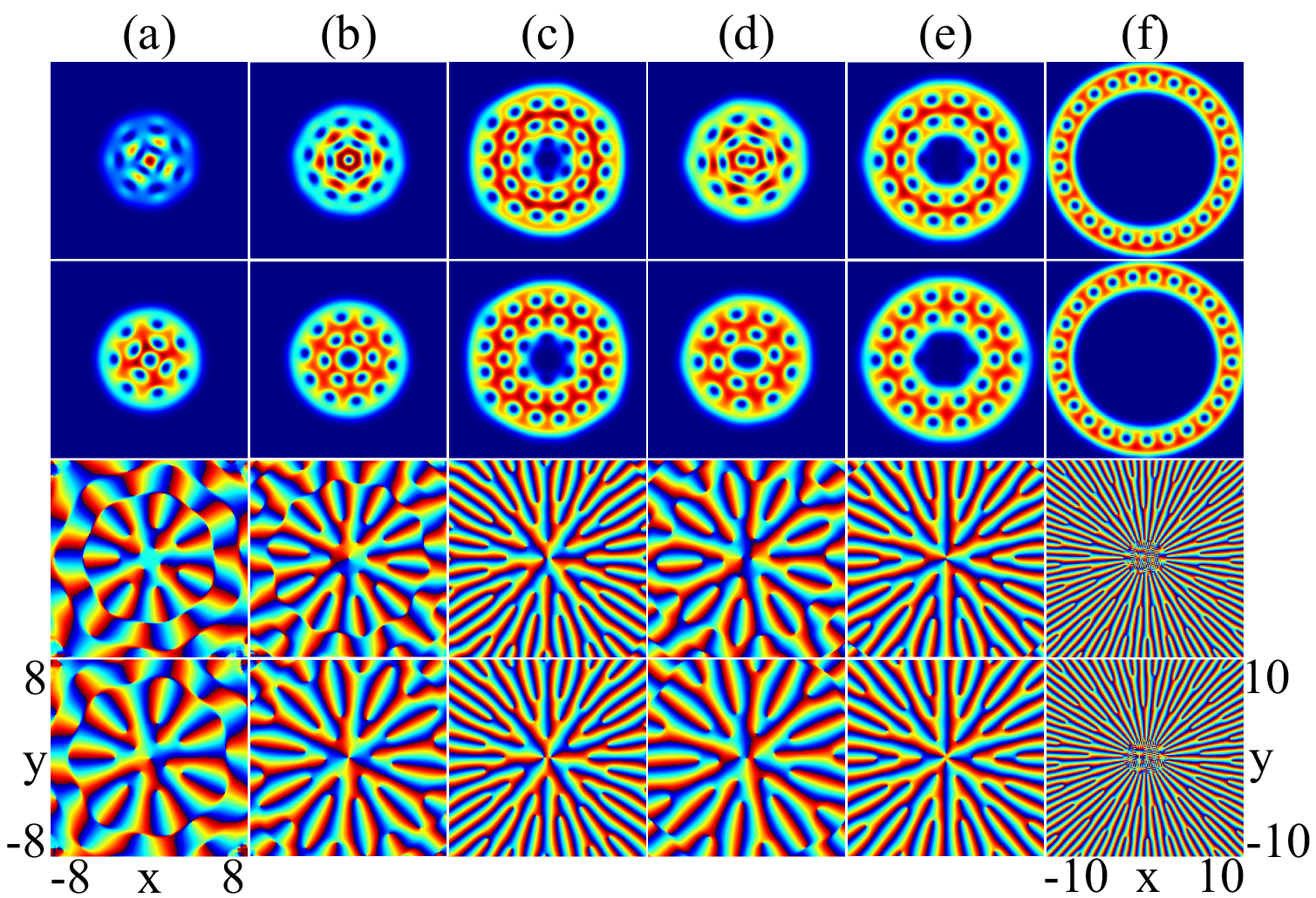}
		\caption{Comparison of harmonic potential and anharmonic potential. (a)-(c) are the density plots of harmonic traps corresponding to the parameters in Figs.~ \ref{Fig-Rotation}(e)-\ref{Fig-Rotation}(g) which are replotted here in (d)-(f) for better comparison.}
		\label{Fig-anharmonic}
	\end{figure}

	\begin{figure*}[t]
		\includegraphics[width=1.0\textwidth]{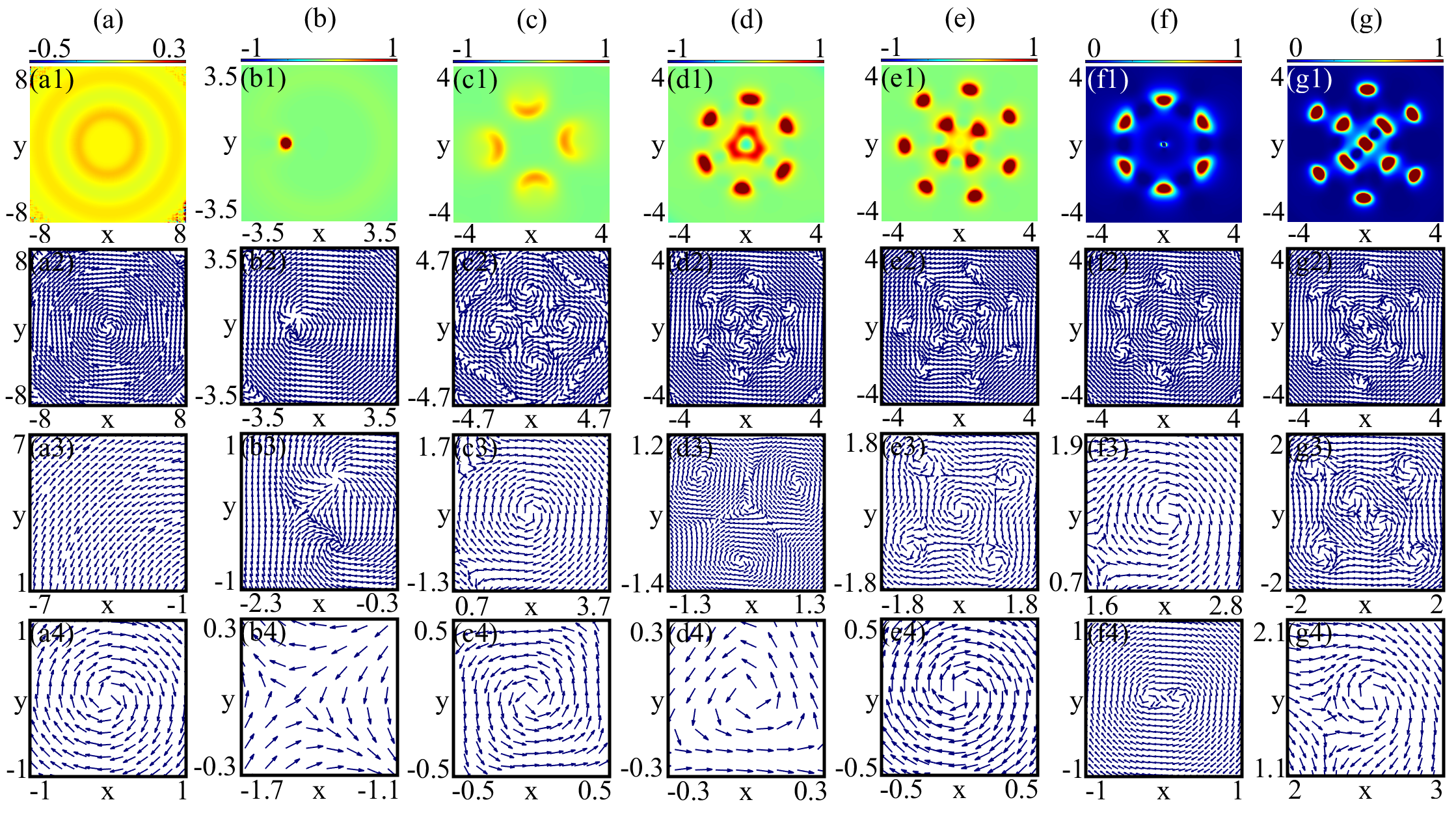}
		\caption{
Exotic topological structures and coexistence states of different topological charge.
(a) Centric vortex surrounded by layers of spin flows.
(b) Compound topological structure of edge topological defect.
(c) Coexisting skyrmions and vortex.
(d) Coexisting skyrmions, half skyrmions and vortex.
(e) Quarter skyrmion and its coexistence with skyrmions.
(f) Coexisting quarter skyrmion and possible three-quarter skyrmions,
(g) Possible fractional skyrmions.
The first row shows the topological charge densities while the other rows show the overall or close-up views of inplane spin textures. Panels (a)-(g) correspond to Figs.~\ref{Fig-DDI}(a,c), \ref{Fig-DDI}(b), \ref{Fig-DDI}(d), \ref{Fig-SOC}(e), \ref{Fig-Rotation}(d), \ref{Fig-SOC}(d) and \ref{Fig-SOC}(f), respectively.
}
		\label{Fig-Topo-Sxy}
	\end{figure*}

\subsection{Effect of potential anharmonicity}

It is also worthwhile to see the effect of the potential anharmonicity as trapping potentials are not ideally harmonic in realistic situation. In Fig.~\ref{Fig-anharmonic} we compare the situation of harmonic potential [panels (a)-(c)] with the anharmonic potential [panels (d)-(f)] under the same parameters of the DDI, SOC and rotation. We see that in anharmonic situation it is easier to generate the density wheel with multiple-layer structure and also quicker to reach the wheel/ring transition mentioned in Section \ref{Sect-Rotation}.  Actually in the multiple-layer wheel and ring-like density distributions, exotic topological quantum states emerge with coexisting vortex, skyrmions and fractional skyrmions, as addressed in the following section.

\section{Spin Texture}\label{Sect-SpinTexture}

In previous sections, we have examined the different effects of the DDI, SOC, rotation and potential anharmonicity, from which we have seen quantum droplets, density defects and giant centric holes. Here we shall show that these states are accompanied with spin structures of vortex without topological charge and skyrmions with integer or fractional topological charge, and they can also coexist to form a compound topological molecule.

\subsection{Exotic spin topological states}

	To better clarify the spin topological structures we present the topological charge density and spin texture. The first row of Fig.~\ref{Fig-Topo-Sxy} shows the topological charge density $q(\bm r)$ as defined in Appendix~\ref{Appendix-S-Q}. An integral of $q(\bm r)$ over the defect area gives the topological charge $Q$, with a finite value of $Q$ characterizing a skyrmion. The overall views of inplane spin textures are presented in the second row of Fig.~\ref{Fig-Topo-Sxy}, while clearer views of the local topological structures of the defects are demonstrated by the close-up plots in the lower rows of Fig.~\ref{Fig-Topo-Sxy}.

\subsubsection{Centric vortex surrounded by layers of spin flows}

Column (a) of Fig.~\ref{Fig-Topo-Sxy} shows the spin topological structure corresponding to the cases with a centric defect in the first component as in Figs.~\ref{Fig-DDI}(c). This case has a vortex with $Q \sim 0$ surrounded by different layers of sin flow as in Fig.~\ref{Fig-Topo-Sxy}(a2). The layer boundaries come from the phase boundaries of the two species components, while the spin flow changes the clockwise or counter-clockwise direction across the spin layer boundaries, as one see in Fig.~\ref{Fig-Topo-Sxy}(a3). The vortex structure can be clearly seen in Fig.~\ref{Fig-Topo-Sxy}(a4) with a clockwise direction.

Actually the centric density defect is not the necessary condition for the appearance of the centric vortex. In reality, the cases of the density disk without the centric density defect, as in Fig.~\ref{Fig-DDI}(a) in repulsive DDI and Figs.~\ref{Fig-DDI}(f)-\ref{Fig-DDI}(h) in attractive DDI, also have similar centric vortex. Such a centric vortex should stem from the rotation, as it disappears in the stripe state when the rotation is turned off in the case of Fig.~\ref{Fig-Rotation}(a) with $\Omega =0$.

\subsubsection{Skyrmions at the quantum droplets and coexistence of skyrmions and vortex}

Column (c) of Fig.~\ref{Fig-Topo-Sxy} shows the case with quantum droplets in Fig.~\ref{Fig-DDI}(d). It turns out that each quantum droplet here carries a skyrmion with a topological charge $Q\sim 1$, as in Fig.~\ref{Fig-Topo-Sxy}(c2) and \ref{Fig-Topo-Sxy}(c3), which is in contrast to the zero topological charge of the centric density defect.
Note here there is also a centric vortex without topological charge as in
Fig.~\ref{Fig-Topo-Sxy}(c4). The centric vortex has a counter-clockwise direction in spin flow, while the skyrmions at the quantum droplet positions are in clockwise directions.  Thus we have a coexistence state of skyrmions and vortex.

\subsubsection{Coexisting skyrmions, half skyrmion and vortex}

Column (d) of Fig.~\ref{Fig-Topo-Sxy} shows the case with two layers of quantum droplets in Fig.~\ref{Fig-SOC}(e).
The outer quantum droplets carry skyrmions with topological charge $Q\sim 1$,
while the inner quantum droplets hold half skyrmions with topological charge $Q\sim 0.5$ instead.
The center density in the first component has a nearly empty region in a shape of three-pointed star but the second component reversely exhibits a density peak in a triangle shape, which hosts a vortex with $Q=0$.
Thus, we have a coexistence state of skyrmions, half skyrmion and vortex.

\subsubsection{Quarter skyrmion and coexistence with skyrmions}

In particular, in column (e) of Fig.~\ref{Fig-Topo-Sxy}, we find an unusual quarter skyrmion with $Q\sim 0.25$. This occurs in case of Fig.~\ref{Fig-Rotation}(d), where we have a centric quantum droplet in the first component. The centric skyrmion in Fig.~\ref{Fig-Topo-Sxy}(e4) manifests a quarter topological charge $Q\sim 0.25$, while the outer skyrmions have $Q\sim 1$. So we have a quarter skyrmion coexisting with skyrmions.

\subsubsection{Quarter skyrmion, three-quarter skyrmions and their coexistence }

We further find possible three-quarter skyrmions with toplogical charge around $Q=0.75$ coexisting with quarter skyrmion, as in column (f) of Fig.~\ref{Fig-Topo-Sxy} which corresponds to Fig.~\ref{Fig-SOC}(d). Here we have a quarter skyrmion at the origin, while the outer skyrmions carry nearly three quarters of topological charge.

\subsubsection{Possible other fractional skyrmions}

In previous cases we have addressed skyrmions, half skyrmions and quarter skyrmions. There may be possibility of other fractional skyrmions. We illustrate a case in column (g) of Fig.~\ref{Fig-Topo-Sxy}. In this case there are three layers of skyrmions as shown in Fig.~\ref{Fig-Topo-Sxy}(g2), a close-up view of the inner two layers is shown in Fig.~\ref{Fig-Topo-Sxy}(g3) and zoom-in view of spin distribution of the middle layer is illustrated in Fig.~\ref{Fig-Topo-Sxy}(g4). The outermost layer is composed of skyrmions with $Q\sim 1$, the middle skyrmions and the innermost ones have topological charges around $Q\sim 0.8$ and $Q\sim 0.6$ respectively. Such a case corresponds to Fig.~\ref{Fig-SOC}(f) where it is a mixture of quantum droplets and stripes due to the increased miscibility by the enhanced SOC.
Although a further study might be needed to clarify whether the fractional $Q$ is intrinsic or results from area overlapping,
similar mixture state of quantum droplets and stripes was encountered in Fig.~\ref{Fig-Rotation}(b), for which we also have fractional skyrmions with $Q\sim 0.33$ at the stripes and $Q\sim 0.64$ at the droplets.

\subsubsection{Compound topological structure of edge topological defect}

In the previous descriptions of spin topological structures, each quantum defect or droplet basically carries a single vortex or skyrmion with some finite topological charge. The edge defect as in Fig.~\ref{Fig-DDI} is distinguished as it induces a compound topological structure. Column (b) of Fig.~\ref{Fig-Topo-Sxy} shows the topological structure of the edge defect. Indeed, as shown in Fig.~\ref{Fig-DDI}(b2), the compound topological structure conatins a N\'{e}el type fractional antiskyrmion with the topological charge $Q\sim 0.6$ right at the defect position [Fig.~\ref{Fig-DDI}(b4)] and a pair of vortices with $Q\sim 0$ and opposite directions beside the fractional antiskyrmion.

These examples of spin texture build up a kaleidoscope of compound topological molecules, indicating a subtle competition and interplay among the DDI, SOC, rotation and anharmonicity. Still, one can get some insights from the correspondence of the spin topological structures to the density topology of the density defect, quantum droplets and density wheel. Indeed, their competition and interplay lead to layer structures which may create a situation to induce fractional skyrmions.

\subsection{Comparison of Rashba-type SOC and Weyl-like SOC}

\begin{figure}
		\includegraphics[width=0.5\textwidth]{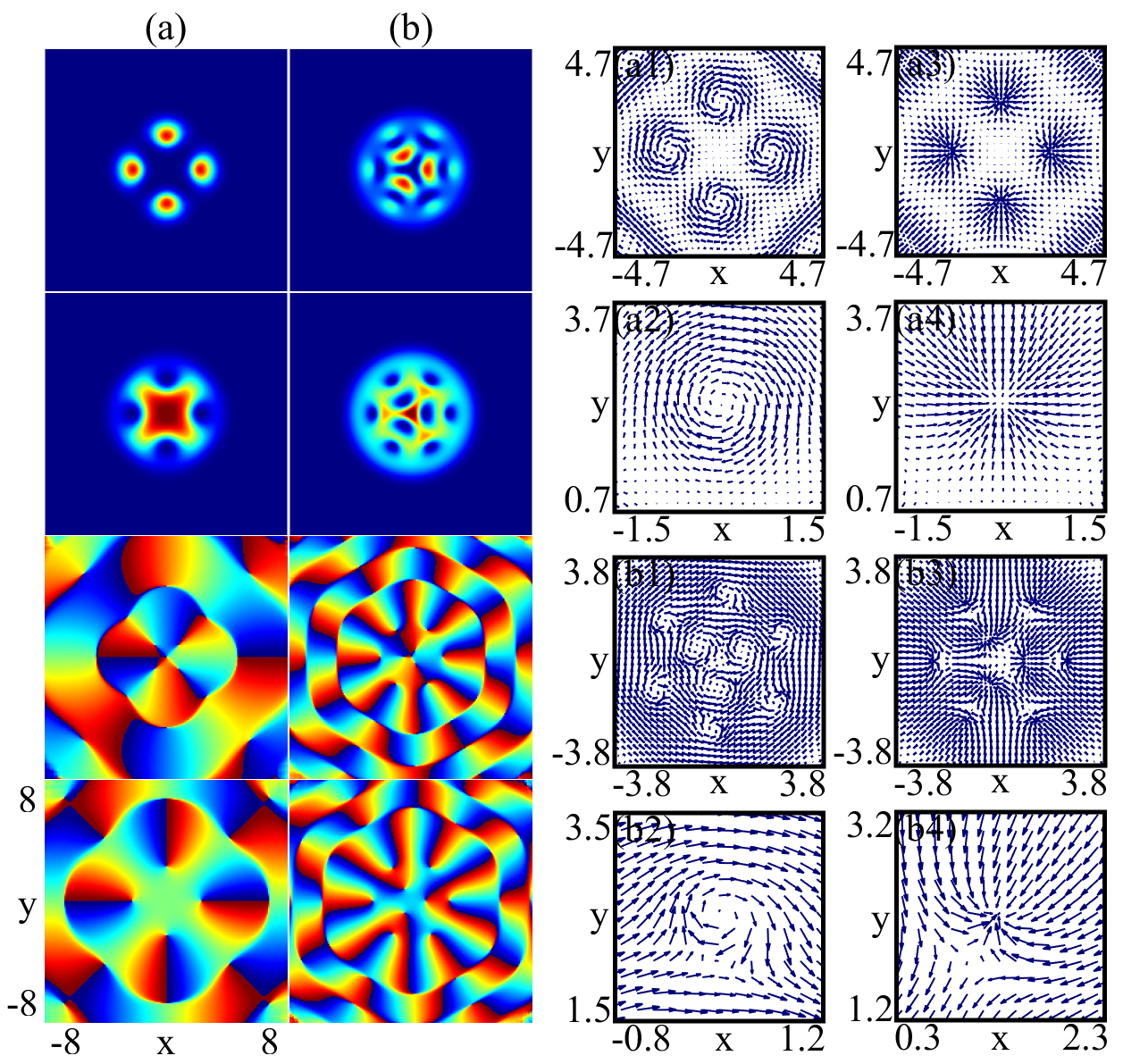}
		\caption{Comparison of Rashba-type SOC and Weyl-like SOC. (a) and (b) show the density plots corresponding to Rashba spin-orbit coupling and Weyl-like spin-orbit coupling in Figs.~\ref{Fig-DDI}(d) and \ref{Fig-SOC}(e), respectively. The third column presents the spin texture for the Weyl-like coupling, while the fourth column shows the spin texture for the Rashba coupling.}
		\label{Fig-Rashba-Weyl}
	\end{figure}

We also compare Rashba-type SOC with the Weyl-like SOC, via replacement of the $k$ terms in Eq.~(\ref{Eq-GP}) by
\begin{eqnarray}
k (\partial_x - i \partial_y) \psi_2,
\\
- k ( \partial_x + i \partial_y) \psi_1 .
\end{eqnarray}
Columns of Figs.~\ref{Fig-Rashba-Weyl}(a) and \ref{Fig-Rashba-Weyl}(b) show the density distribution and phase structure of the two components in Rashba-type SOC, with the same parameters of the DDI, rotation and potential corresponding to the Weyl SOC cases in Figs.~\ref{Fig-DDI}(d) and \ref{Fig-SOC}(e). We see that the Rashba-type SOC and Weyl-like SOC share the same density distributions.

On the other hand, we compare the spin textures. The spin textures in the Rashba-type SOC are illustrated in Figs.~\ref{Fig-Rashba-Weyl}(a3) and \ref{Fig-Rashba-Weyl}(b3), with corresponding close-up view in  Figs.~\ref{Fig-Rashba-Weyl}(a4)\ref{Fig-Rashba-Weyl}(b4). For a convenient comparison, the spin textures in the Weyl-type SOC are replotted in Figs.~\ref{Fig-Rashba-Weyl}(a1) and \ref{Fig-Rashba-Weyl}(b1), with corresponding close-up view in Figs.~\ref{Fig-Rashba-Weyl}(a2) and \ref{Fig-Rashba-Weyl}(b2). We find that the spin textures in the Weyl-type SOC and Rashba-like SOC have perpendicular in-plane spin directions. As a result, when the spin flow is counter-clockwise (clockwise) in the Weyl-type SOC, the spin flow in the Rashba-type SOC will be inward (outward). A Block-type skyrmion in the Weyl-type SOC will become a N\'{e}el-type skyrmion the Rashba-type SOC, as more clearly demonstrated in Figs.~\ref{Fig-Rashba-Weyl}(a2), \ref{Fig-Rashba-Weyl}(a4), ~\ref{Fig-Rashba-Weyl}(b2) and \ref{Fig-Rashba-Weyl}(b4).

\section{Conclusions}\label{Conclusion}

We have investigated the ground state of two-component dipolar BECs with the Weyl-like SOC in a rotating 2D anharmonic potential trap. A systematical study has been  carried out on the effects the DDI, SOC, rotation and anharmonicity in their interplay, with their influences on the distributions of density, phase and spin texture. We have revealed a kaleidoscope of exotic quantum states, with transitions of density topology,  critical behavior around a revealed collapse point and a bunch of coexisting states of spin structures with integer and fractional topological charges.

We have shown that the repulsive DDI leads to density variations of the first component from disk, edge defect, centric defect to quantum droplets, while the density of the other component transit from a ring shape, to cross shape, pentagram, mutli-pointed stars, with increasing number of holes or defects on the margin. The variations actually experiences transitions of density topology. In attractive DDI we have found a quantum phase transition after which the density collapses into a sharp spike. In particular, the size dependence on the attractive DDI strength manifests a criticality. During the variation of the DDI either repulsive or attractive, the ground state exhibits phase separation in the sense that the two components are basically separated in embedding or surrounding each other.

The SOC also not only induces transitions in the density topology similar to the DDI, but also increases the layers of density defects and quantum droplets. Concerning the phase separation the SOC plays an opposite role relative to the DDI, by increasing the miscibility of the two components. And a large SOC also expands size of the density wheel.

The rotation first converts into the stripes state into layer structure of density defects and quantum droplets, then beyond some critical strength of rotation the density wheel is turned to rings with density holes or defects. The rotation also increases the miscibility of the two components and finally leads similar position distribution of holes or defects in the two components. Compared with the SOC, the rotation enlarges the size of the density wheel or ring, but finally reduces the topological layers.

We have also seen that in anharmonic situation it is easier to generate the density wheel with multiple layer structure, and quicker as well to reach the wheel/ring transition.

On the other hand, in the spin texture we have unveiled various exotic spin topological states.
Indeed, we have shown centric vortex surrounded by layers of spin flows in opposite directions, compound topological structure of edge topological defect, coexisting skyrmions and vortex, coexisting skyrmions, half skyrmion and vortex, quarter skyrmion and its coexistence with skyrmions, coexisting quarter skyrmion and three-quarter skyrmions, and also possibility of other fractional skyrmions.

Finally we have compared the Rashba-type SOC and Weyl-like SOC. We find that these two types of SOCs yield same density distributions  but have perpendicular in-plane spin directions. As a result, when the spin flow is counter-clockwise or clockwise in the Weyl-type SOC, the spin flow in the Rashba-type SOC will be inwards or outwards. Particularly, a Block-type skyrmion in the Weyl-type SOC will become a N\'{e}el-type skyrmion the Rashba-type SOC.

The emergence of these exotic quantum states are results of the competition and interplay of the DDI, SOC, rotation and anharmonicity. Our study implies
that one can manipulate both the topology of density distribution and the topological structure of spin distribution via tuning these parameters,
which is feasible due to the high controllability of cold atom systems~\cite{Li2012PRL,LinRashbaBECExp2011,LinRashbaBECExp2013Review,wu2016realizeSOC,Campbell2016RealizeSpin1SOC,GongMing2019BECSOC,Anderson2013PRLmagnGenerateSOC}.
The abundant variations of the topological structures and particularly the revealed critical behavior may also provide quantum resources for potential applications, e.g., in quantum metrology~\cite{RamsPRX2018,Garbe2020,Montenegro2021-Metrology,Ilias2022-Metrology,Ying2022-Metrology,Hotter2024-Metrology,Ying-Topo-JC-nonHermitian-Fisher,
Ying-g2hz-QFI-2024,Ying-g1g2hz-QFI-2025,Ying2025g2A4,Ying2025TwoPhotonStark} which exploits the variation resource of the ground state for parameter measurements~\cite{Cramer-Rao-bound,RamsPRX2018,Ying-gC-by-QFI-2024}. Such an application deserves a special discussion which we shall address elsewhere.

	\begin{acknowledgments}
			
This work was supported by the National Natural Science Foundation of China
(Grants No. 12474358, No. 11974151, and No. 12247101).

	\end{acknowledgments}

\appendix\bigskip

\section{Dimensionless formulism of GP equation and numerical detail}
\label{Apendix-method}

\subsection{Convolution treatment for DDI}

For numerical simulations, the DDI can be efficiently handled in momentum space by employing the convolution theorem via Fourier transformation, which is expressed as
\begin{align}
\Phi({\bm{\mathrm{r}}},t)=\frac{\mu_{0} \mu^{2} m}{3\sqrt{2\pi} \hbar^{2} a_{z}}
\mathcal{F}_{2\mathrm{D}}^{-1}[\tilde{n}({\bm{\mathrm{k}}},t) F(\bm{\mathrm{k}} a_{z}/\sqrt{2})].
\end{align}

Here, $\mathcal{F}_{2\mathrm{D}}$ denotes the 2D Fourier transform operator, such that the Fourier-transformed density is given by $\tilde{n}({\bm{\mathrm{k}}},t)=\mathcal{F}_{2D}[n({\bm{\mathrm{r}}},t)]$. The dipole-dipole interaction (DDI) in momentum space for a quasi-2D system is characterized by the function $\bm{\mathrm{q}}\equiv \bm{\mathrm{k}}\alpha_{z}/\sqrt{2}$. This interaction comprises contributions associated with dipole polarization components either perpendicular or parallel to the tilt direction. Specifically, ${\cal{F}}(\bm{\mathrm{q}})=\cos^{2}(\alpha)F_{\perp}(\bm{\mathrm{q}})+\sin^{2}(\alpha)F_{\|}(\bm{\mathrm{q}})$,
	with $\alpha$ representing the angle between the dipole orientation vector
	$\hat{d}$ and the $z$-axis. Here
	
	\begin{eqnarray}	
	&&{\cal F}_{\perp}(\bm{\mathrm{q}})=2-3\sqrt{\pi}q e^{q^{2}}\mathrm{erfc}(q),\\
	&&{\cal F}_{\parallel}(\bm{\mathrm{q}})=-1+3\sqrt{\pi}\left(\frac{q_{d}^{2}}{q}\right)e^{q^{2}}\mathrm{erfc}(q),
	\end{eqnarray}
	where $q_{d}$ denotes the projection of the wave vector along the in-plane component of the dipole moment, and $\mathrm{erfc}(q)$ is the complementary error function.
	In the special case where the dipoles are aligned along the $z$-axis ($\alpha=0$), the interaction reduces to ${\cal F}(\bm{\mathrm{q}})={\cal F}_{\perp}(\bm{\mathrm{q}})$ and the dipoles are aligned along the $x$-axis  ($\alpha=\pi/2$), yielding $F(\bm{\mathrm{q}})=F_{\parallel}(\bm{\mathrm{q}})$.

\subsection{Dimensionless representation for the GP equation }

To facilitate numerical computations and simulations, we adopt the following renormalized notations $\tilde{r}=r/a_{0},
	\tilde{t}=\omega_{\perp}t,
	\tilde{V}(r)=V(r)/\hbar\omega_{\perp}, \tilde{\phi}_{1}=\phi_{1}/\hbar\omega_{\perp},
	\tilde{\psi}_{j}=\psi_{j}a_{0}/\sqrt{N}(j=1,2),
	\beta_{j j}=g_{j}N m/\hbar^{2}(j=1,2),
	\beta_{12}=g_{12}N m/\hbar^{2}.$
Then we obtain the dimensionless 2D coupled GP equations of the system
	\begin{eqnarray}
i\partial_{t}\psi_{1} &=& \left(-\frac{1}{2}\nabla^{2} + V + \beta_{11}\left|\psi_{1}\right|^{2} + \beta_{12}\left|\psi_{2}\right|^{2} - \Omega L_{z} + \Phi\right)\psi_{1} \notag \\
		&&\quad - k(\partial_{y} - i\partial_{x})\psi_{2}, \notag \\
i\partial_{t}\psi_{2} &=& \left(-\frac{1}{2}\nabla^{2} + V + \beta_{12}\left|\psi_{1}\right|^{2} + \beta_{22}\left|\psi_{2}\right|^{2} - \Omega L_{z}\right)\psi_{2} \notag \\
		&&\quad + k(\partial_{y} - i\partial_{x})\psi_{1}.
	\end{eqnarray}
Here for simplicity we have dropped the tildes over renormalized parameters and the expression for the 2D potential becomes
	$V=\frac{1}{2}r^{2}+A\mathrm{e}^{-n^{2} r^{2}}$.
When the value of $n$ is small the potential can be approximated as $V=A+\frac{1}{2}(1-2A^2n^2)r^2+\frac{1}{2}An^4r^4$.

	To characterize the relative strength of the dipolar and s-wave interactions, a dimensionless parameter $\varepsilon_{dd}$ is introduced. Indeed, the three-dimensional dipolar length is defined as $a_{dd}^{3D}=\mu_0\mu^2m/12\pi\hbar^2$. For the quasi-2D system, the effective dipolar length becomes $a_{dd}^{2D}=4\pi C_{dd}/3\sqrt{2\pi}a_{z}=\mu_0\mu^2m/3\sqrt{2\pi}\hbar^2a_{z}$. The
	dipolar coupling coefficient is given by
\begin{equation}
\varepsilon_{dd}=a_{dd}/a_{s}=\mu_0\mu^2m/3\sqrt{2\pi}\hbar^2a_{z}/a_{s},
\end{equation}
where $a_s$ denotes the s-wave scattering length. Assuming the dipoles are aligned side by side along the $z$-axis, the DDI becomes isotropic, manifesting as either repulsion or attraction, and can be approximated as an effective contact interaction. Consequently, the DDI may be expressed in the form $\phi_1=\beta_{11}\varepsilon_{dd}|\psi_1|^2$, where $\varepsilon_{dd}>0$
	corresponds to repulsion and $\varepsilon_{dd}<0$ indicates attraction when $a_s>0$. The total interaction coefficient for component $1$ is given by $(1+\varepsilon_{dd})\beta_{11}$. By adjusting $\varepsilon_{dd}$, the SOC strength $k$ and the interaction parameters $\beta_{11},\beta_{22}$ and $\beta_{12}$, a variety of ground-state phases can be achieved as addressed in Section \ref{Sect-Effects}.

\subsection{Parameters setting in numerics}

	Due to the complexity of the system, obtaining analytical solutions is not feasible. To investigate the ground state properties of the system, we numerically solve the coupled GP equations (4) using the imaginary-time propagation method. In our study, we systematically explore the combined effects of the DDI, Weyl-like SOC and rotation frequency on the ground-state BECs confined in the anharmonic trap. As an illustration, we fix the interaction parameters $\beta_{11}=50,\beta_{22}=80,\beta_{12}=200,$ and set the parameter of anharmonic trap $A=25,l=0.08$. Variation in the parameters of the DDI strength $\varepsilon_{dd}$, SOC strength $k$ or rotation frequency $\Omega$, will lead to the emergence of rich quantum states.

\section{Pseudospin and topological charge}\label{Appendix-S-Q}
	
	To describe the spatial distribution of the topological structure, we utilize a nonlinear Sigma model~\cite{Kasamatsu2005PRA} in which a normalized complex-valued spinor $\chi=[\chi_1,\chi_2]^T$ with $|\chi_1|^2+|\chi_2|^2=1$ is introduced. This approach maps the system to a magnetic system by employing a pseudospin representation of the order parameter, enabling a deeper understanding of systems with internal degrees of freedom. For instance, two-component BECs can be treated as spin-1/2 systems, where $\psi_1(\psi_2)$
	corresponds to the spin-up (spin-down) component of the spinor BEC. The total density of the system is expressed as $\rho=|\psi_1|^2+|\psi_2|^2$ and the corresponding wave functions are $\psi_1=\sqrt{\rho}\chi_1$ and $\psi_2=\sqrt{\rho}\chi_2$ .
	In the pseudospin framework, the spin density is defined as $\mathbf{S}=\chi^{\dagger}\sigma\chi$, where ${\boldsymbol{\sigma}}=\left(\sigma_{x},\sigma_{y},\sigma_{z}\right)$ are the Pauli matrices. The spin components are given by~\cite{Mizushima2004PRA,Kasamatsu2005PRA}
	\begin{eqnarray}
		S_{x}&=&\chi_{1}^{*}\chi_{2}+\chi_{2}^{*}\chi_{1},\notag \\
		S_{y}&=&i(\chi_{2}^{*}\chi_{1}-\chi_{1}^{*}\chi_{2}),\notag \\
        S_{z}&=&|\chi_{1}|^{2}-|\chi_{2}|^{2},
	\end{eqnarray}
with $|\mathbf{S}|^{2}=S_{x}^{2}+S_{y}^{2}+S_{z}^{2}=1$. The spatial distribution of the system's topological structure is characterized by the topological charge density
\begin{equation}
q({\mathbf r})={\frac{1}{4\pi}}\,{\mathbf S}\cdot\left({\frac{\partial{\mathbf S}}{\partial x}}\times\,{\frac{\partial{\mathbf S}}{\partial y}}\right),
\end{equation}
and the total topological charge is $Q=\int q({\mathbf r})d x d y$. This formalism provides a mapping between the internal degrees of freedom of the system and a magnetic system, facilitating an understanding of its topological properties.

	\bibliography{ref-DDI}

\end{document}